\documentclass[onecolumn,secnumarabic,amssymb, nobibnotes, aps, prd,12pt]{revtex4-1}

\usepackage{wrapfig}
\usepackage{dsfont}
\usepackage{amscd}
\usepackage{amsmath}
\usepackage{amssymb}
\usepackage{graphicx}
\usepackage{bm}
\usepackage{newtxtext}
\usepackage{booktabs,caption} 

\usepackage{comment}

\usepackage{subcaption}
\usepackage{sidecap}
\usepackage{grffile}
\usepackage{xcolor}
\usepackage{scalerel}

\definecolor{rred}{rgb}{0.7,0,0.1}
\definecolor{greenrb}{rgb}{0.2,0.6,0.2}

\def\bi{\begin{itemize}}
\def\ei{\end{itemize}}

\def\bea{\begin{equation} \begin{aligned}}
\def\eea{\end{aligned} \end{equation}}
\def\beas{\begin{equation*} \begin{aligned}}
\def\eeas{\end{aligned} \end{equation*}}
\def\bes{\begin{equation*}}
\def\ees{\end{equation*}}

\def\be{\begin{equation}}
\def\ee{\end{equation}}

\newcommand{\beq}{\begin{equation}}
\newcommand{\eeq}{\end{equation}}
\newcommand{\beqs}{\begin{subequations}}
\newcommand{\eeqs}{\end{subequations}}
\newcommand{\balign}{\begin{align}}
\newcommand{\ealign}{\end{align}}

\begin{document}

\title{Latitudinal storm track shift in a reduced two-level model of the atmosphere}    
 \author{Melanie Kobras}
 \affiliation{Department of Mathematics and Statistics, University of Reading, Reading, RG6 6AX, UK}
 \affiliation{Centre for the Mathematics of Planet Earth, University of Reading, Reading, RG6 6AX, UK}
 \author{Maarten H. P. Ambaum}
 \affiliation{Department of Meteorology, University of Reading, Reading, RG6 6AX, UK}
  \author{Valerio Lucarini}
\email{Corresponding author. Email address: \texttt{v.lucarini@reading.ac.uk}}\affiliation{Department of Mathematics and Statistics, University of Reading, Reading, RG6 6AX, UK}
 \affiliation{Centre for the Mathematics of Planet Earth, University of Reading, Reading, RG6 6AX, UK}
 
\date{\today}

\begin{abstract}

The eddy-driven jet stream and storm tracks in the mid-latitude atmosphere are known to shift in latitude on various timescales, but the physical processes that cause these shifts are still unclear. In this study, we introduce a minimal dynamical system derived from the classical Phillips two-level model with the goal of elucidating the essential mechanisms responsible for the interaction between eddies and mean flow. Specifically, we aim to understand the link between the structure of the eddies and the shift of the latitudinal maximum of the zonal flow. By varying the horizontal shape of the eddies, we find three distinct dynamical regimes whose occurrence depends on the intensity of the external baroclinic forcing: a purely zonal flow, a barotropic eddy regime with net poleward momentum flux, and a baroclinic eddy regime with both net poleward momentum and temperature flux. For weak baroclinic forcing, the classical zonal flow solution with latitudinal maximum at the centre of the beta-channel is found. For strong forcing, if eddies are southwest-northeast tilted and zonally elongated, the system is in the baroclinic eddy regime, resulting in a poleward shift of the jet. The intermediate barotropic eddy regime also features a poleward shifted jet, yet with eddies structurally distinct from the baroclinic regime. Changing the parameters yields transitions between the regimes that can be either continuous or discontinuous in terms of the properties of the atmosphere. The findings of this study also provide insights into the properties of the storm track change between the jet entrance and jet exit regions of the North Atlantic. 

\end{abstract}

\maketitle    

\section{Introduction}
The eddy-driven jet stream and associated storm track is the main location of extreme weather events in the mid-latitude atmosphere and has a profound impact on the general circulation of the atmosphere and oceans by transporting heat, momentum and moisture from the equator towards the poles. The strength and location of the jet stream vary on a vast range of 
time scales 
(\cite{Holton2013Dynamic}). 
In order to explain these variations and eventually be able to predict them, it is important to understand the driving mechanisms.

Numerous studies quantifying and describing the impact of the eddies on the zonal flow and its location evolved around the work of \cite{Hoskins1983TheShape} who developed a diagnostic time-averaged quantity, which they called E-vector, in order to assess the three-dimensional momentum convergence and the influence of the eddies on the mean flow. They also showed the important role played by the anisotropy of the eddies - and specifically by their tilt and aspect ratio.

Using data for the Northern Hemisphere 1979-80 winter from the European Centre for Medium-Range Weather Forecasts (ECMWF), the diagnostic showed that in the entrance region of the storm track non-tilted eddies with similar zonal and meridional wavenumber transport momentum to lower levels and decrease the vertical shear in the zonal flow, and therefore enhance the barotropic nature of the westerly flow near the entrance and in the middle sector. Instead, in the jet exit region, zonally elongated southwest-northeast tilted eddies decrease the kinetic energy of the mean flow. 

The analysis by \cite{Woollings2010Variability}, based on the 40-year ECMWF re-analysis (ERA-40) data, suggests that there are three preferred latitudinal positions of the North Atlantic eddy-driven jet stream in winter: south, middle, and north. Using a nonlinear oscillator relationship between the meridional temperature gradient (baroclinicity) and the meridional eddy heat flux, 
\cite{Novak2015TheLife} found that high upstream eddy heat flux tends to deflect the jet northward, whereas low eddy heat flux is associated with a more southward deflected jet. We remind that the position of the jet changes also on much longer time scales. The observed slow poleward shift of the mid-latitude storm track \citep{Fyfe2003Extratropical} is projected to continue as the climate warms \citep{Yin2005Aconstistant}, as a result of changes in the baroclinicity and in the static stability of  the mid-latitude atmosphere as well as in the height of the tropopause \citep{Lorenz2007,Frierson2008,Lu2010,Butler2010}. 
However, a generally accepted theory for the dynamical mechanisms driving observed and modelled shifts is still absent; see discussion in \cite{Thompson2000,Kidson2010}. 

Instead of using re-analysis data or large model simulations, this study aims at 
asking the question of whether it is possible to describe at least at a qualitative level the shifts of the jet as behaviour in a severely truncated model of the atmosphere. Our objective is not to create a realistic model of atmospheric dynamics but rather to capture the essential nonlinear processes responsible for the latitudinal vacillation of the jet by distilling the minimal components needed to observe such behaviour. 

Based on Phillips' two-level quasi-geostrophic model on the $\beta$-plane \cite{Phillips1956Thegeneral}, arguably the simplest model of the dynamics of the mid-latitude atmosphere that can incorporate both barotropic and baroclinic processes, with external forcing associated with diabatic heating, \cite{Kobras2021Eddy} developed a set of ordinary differential equations that are able to provide a minimal yet meaningful model of the interaction between mean flow and eddy activity. In that work, the eddies are assumed to have no east-west tilt. Hence, eddy momentum fluxes were not represented and the maximum of the mean flow did not vary in latitude.

The novelty of the present paper is that we allow a tilt in the eddies and redefine the shape of the mean flow in order to allow for variations of the latitudinal maximum of the zonal wind. In accordance with that, we choose a southwest-northeast tilted eddy shape to allow nonlinear interactions in form of northward eddy momentum fluxes. These changes yield substantially richer dynamics as compared to \cite{Kobras2021Eddy}. Apart from a purely zonal regime, our new model exhibits competing barotropic and baroclinic eddy regimes with different impacts on the latitudinal maximum of the mean flow, or, namely, the position of the jet. The derivation of our minimal model differs substantially from classical modal expansion as presented in e.g. \cite{Koo2002,Thompson1987}.

This paper is structured as follows. In section \ref{sec:model_equations} we briefly review the model derived in \cite{Kobras2021Eddy}, followed by above mentioned definitions of the mean flow and eddy shape, and the derivation of evolution equations for the flow amplitudes defining our model. In section \ref{sec:zonal_flow} the steady state and stability thereof are determined and the model is compared to the non-tilted eddy case in \cite{Kobras2021Eddy}. The complete dynamics of the system for tilted eddies is described in section \ref{sec:eddy_regimes}, and section \ref{sec:discussion} summarises the results and discusses them in light of the current literature.

\section{The model equations}\label{sec:model_equations}
We use the two-level quasi-geostrophic model in pressure coordinates of \cite{Phillips1956Thegeneral} and follow the derivation in \cite{Kobras2021Eddy} to arrive at evolution equations for the mean zonal wind $\overline{u}_m$ and shear $\overline{u}_T$: 
\begin{subequations}
\begin{align}
    &\frac{\partial \overline{u}_m}{\partial t} -\frac{1}{l^2}  \frac{\partial^3}{\partial y^3}\overline{v'_m u'_m}   -\frac{1}{l^2}\frac{\partial^3}{\partial y^3}\overline{v'_T u'_T} = -l^2 A\overline{u}_m - \frac{\kappa}{2}(\overline{u}_m-2\overline{u}_T), \label{eq:lat1a}\\
    &\frac{\partial \overline{u}_T}{\partial t} -  \frac{1}{l^2+\lambda_R^2} \frac{\partial^3}{\partial y^3}\overline{v'_T u'_m} - \frac{1}{l^2+\lambda_R^2} \frac{\partial^3}{\partial y^3}\overline{v'_m u'_T} -\frac{\lambda^2_R}{l^2+\lambda_R^2}\frac{\partial^2}{\partial y^2}\overline{v'_m\psi'_T}\label{eq:lat1b} \\
    &\qquad \qquad \qquad \qquad = -l^2 A\overline{u}_T + \frac{l^2}{l^2+\lambda_R^2}\frac{\kappa}{2}(\overline{u}_m-2\overline{u}_T) + \frac{\lambda^2_R}{l^2+\lambda_R^2}\frac{2RH}{f_0c_pW}, \nonumber
\end{align}
\end{subequations}
where $\boldsymbol{V}=(u,v)$ is the horizontal velocity vector. \textcolor{black} {Additionally, for any field $F$, we have that $F_m(x,y,t)=1/2(F_1(x,y,t)+F_2(x,y,t))$, $F_T(x,y,t)=F_1(x,y,t)-F_2(x,y,t)$ , where the subscripts $_1$ and $_2$ refer to the 250 and 750 hPa levels respectively}, the overbar denotes the zonal mean \textcolor{black}{$\overline{F}(y,t)=(1/L)\int_0^L {\rm d}x\,F(x,y,t)$, and the prime denotes the deviation from such a mean as $F'(x,y,t)=F(x,y,t)-\overline{F}(y,t)$ for any field $F$.} 
These equations are defined on the spatial domain $(x,y)\in[0,L]\times[0,W]$, where $L$ and $W$ are the length and width of the $\beta$-plane channel with $f=f_0+\beta y.$

The parameters $A$ and $\kappa$ describe the eddy diffusion and the surface friction diffusion, hence controlling the dissipation processes that remove energy from the system. The parameter $c_p$ is the isobaric specific heat capacity of dry air and $R$ is the specific gas constant for dry air. 


\textcolor{black}{
The profile of the diabatic heating is linear in $y$, and antisymmetric with respect to the central latitude, so that $H$ is the mean rate of heating per unit mass for $y\in[0,W/2]$ (or cooling for $y\in[W/2,W]$).} Diabatic forcing of this form acts as a baroclinic forcing increases the mean available potential energy of the system by creating a meridional temperature gradient \citep{Lorenz1955Available,Lorenz1967Thenature}. The eddies resulting from the breakdown of the stability of the zonal flow solution transport momentum and heat and act as negative feedback of the system \citep{Lucarini2007Parametric}. 
\textcolor{black}{Finally, the parameter $l$ is the latitudinal wavenumber, and } 
the parameter $\lambda_R^2=2p_2f^2_0/(\delta p^2\overline{\Gamma}_2R)$ is the inverse square of the Rossby radius of deformation,  where $\delta p$ is the pressure difference between levels 1 and 3 and the basic state static stability is defined as $\overline{\Gamma}=RT/pc_{\rm p}-d T/d p$ with temperature $T$.

The mean zonal wind and shear are defined as
\begin{align}
    \overline{u}_{m,T}(t)={\overline{U}_{m,T}}(t)\sin{ly}-{\overline{L}_{m,T}}(t)\cos{ly},\label{eq:lat2a}
\end{align}
with time-dependent amplitudes ${\overline{U}_{m,T}}(t)$ determining the flow speed and (positive) ${\overline{L}_{m,T}}(t)$ and where the longitudinal wavenumber is $l=\pi/W$. 

The latitudinal position $y_{max}$ of the maximum of the zonal mean barotropic flow and zonal shear is given by  $\tan(ly_\text{max})=-{\overline{U}_{T}}/{\overline{L}_{T}}$
and  $\tan(ly_\text{max})=-{\overline{U}_{m}}/{\overline{L}_{m}}$, respectively. Instead, the latitudinal position $y_{max}$ of the maximum of the zonal flow at the two levels is given by
\begin{equation}
\tan(ly_\text{max})=-\left({\overline{U}_{m}} \pm {\overline{U}_{T}}\right)/\left({\overline{L}_{m}} \pm {\overline{L}_{T}}\right)
\end{equation}
with the plus sign corresponding to the upper level jet and the minus sign to the lower level jet, respectively. 


\begin{figure}
\centering
\begin{subfigure}{.99\textwidth}
  \centering
  \includegraphics[width=1\linewidth]{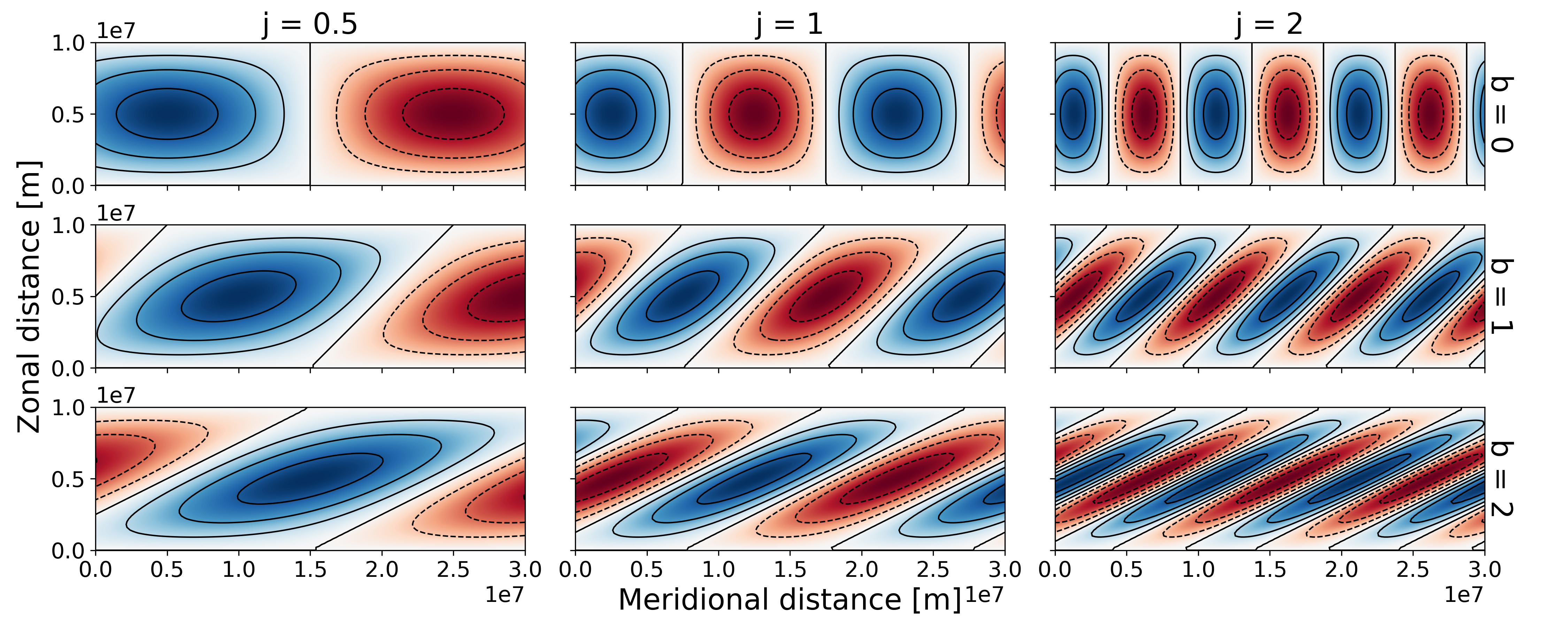}
\end{subfigure}
\caption{Eddy streamfunction shape for different values of the tilt $b$ and the aspect ratio $j$; blue denotes positive, red denotes negative values.}
\label{fig:tilted}
\end{figure}

Based on findings by \cite{Hoskins1983TheShape} and \cite{Orlanski1998Poleward}, showing that during their life cycle, the mid-latitude eddies in the storm track associated with the eddy-driven jet stream evolve to become meridionally elongated and southwest-northeast tilted, the eddy component of the barotropic and baroclinic streamfunction is defined to be of the shape
\begin{equation}
    \psi'_{m,T}(x,y,t)={A_{m,T}}(t)\sin k(x-by)\sin ly+{B_{m,T}}(t)\cos k(x-by)\sin ly, \label{eq:lat3}
\end{equation}
where ${A_{m,T}}(t)$ and ${B_{m,T}}(t)$ are time-dependent amplitudes, the longitudinal wavenumber $l=\pi/W$ as above, and $k$ is a multiple of $l$ defined as $k=jl=j\pi/W$ where $j$ is a parameter describing the aspect ratio of the eddies, that is --- without loss of generality --- assumed to be a non-negative number. In the following, eddies with equal zonal and meridional wavenumber and thus aspect ratio $j=1$ are referred to as being circular. The parameter $b$ describes the slope of the tilt, see figure \ref{fig:tilted}.

In comparison to \cite{Kobras2021Eddy}, such a shape of the eddy streamfunction components does not lead to a cancellation of the eddy product terms in equations (\ref{eq:lat1a}) and (\ref{eq:lat1b}). This is due to the eddies being not meridionally symmetric and therefore the system exhibits non-zero momentum and so-called mixed momentum fluxes defined, respectively, by
\begin{subequations}
\begin{align}
    \overline{u'_{m} v'_{m}}+\overline{u'_{T} v'_{T}} &= \frac{1}{2}j^2 l^2b\left({A^2_{m}}+ {B^2_{m}}+{A^2_{T}}+ {B^2_{T}}\right) \sin^2 ly, \label{eq:mom_flux}\\
    \overline{u'_mv'_T}+\overline{u'_Tv'_m} &= j^2l^2b\left({A_mA_T}+{B_mB_T}\right)\sin^2 ly. \label{eq:mix_flux}
\end{align}
\end{subequations}
We remark that the sum of these fluxes is the momentum flux at the upper level 1 of the two-level model and the difference is the momentum flux at the lower level 3. As can be seen from these equations, the momentum fluxes are proportional to the tilt parameter $b$ and circular eddies ($b=0$) do not produce momentum fluxes.

The net poleward eddy heat flux 
can be calculated as
\begin{align}
    \overline{v'_m T'_2} &= \frac{jlf_0 }{R}\left({A_mB_T}-{A_TB_m}\right)\sin^2 ly, \label{eq:temp_flux}
\end{align}
where the temperature at level 2 is related to the streamfunction $\psi_T$ by the hydrostatic balance equation, and where we have neglected the (physically important) factor involving the heat capacity. The meridional shape of the fluxes closely resembles the zonally averaged reanalysis data as shown in \cite{Kallberg2005ERA-40}. In contrast to the momentum fluxes, the poleward heat flux is not proportional to the eddy tilt $b$ and therefore possibly non-vanishing for non-tilted eddies ($b=0$) as well. 

Next, we derive time evolution equations for the mean flow amplitudes ${\overline{U}_{m,T}}(t)$ and ${\overline{L}_{m,T}}(t)$, and for the eddy streamfunction amplitudes ${A_{m,T}}(t)$ and ${B_{m,T}}(t)$. The former can be obtained by first replacing the mean zonal wind and shear in equations (\ref{eq:lat1a}) and (\ref{eq:lat1b}) by its definitions (\ref{eq:lat2a}), additionally calculating the zonally averaged eddy momentum and heat fluxes occurring in these equations by replacing the eddy variables by their definitions (\ref{eq:lat3}) and projecting both equations onto the two meridional modes of $\overline{u}_m$ and $\overline{u}_T$. This is done by multiplying the result by either $\sin{ly}$ or $\cos{ly}$ and taking the meridional average from the equator to the pole.

To obtain evolution equations for the eddy amplitudes ${A_{m,T}}$ and ${B_{m,T}}$ we project the non-averaged evolution equations for the barotropic and baroclinic vorticities derived in \cite{Kobras2021Eddy} onto the two zonal modes ($\sin{k(x-by)}$ and $\cos{k(x-by)}$) and the meridional mode ($\sin{ly}$) of the barotropic and baroclinic streamfunction (\ref{eq:lat3}), yielding four equations.

Finally, the eight evolution equations for the mean flow and eddy amplitudes are
{\footnotesize
\begin{subequations}
\begin{align}
    \frac{d}{d t}{\overline{U}_m} =& -l^2A{\overline{U}_m}- \frac{\kappa}{2}\left({\overline{U}_m}-2{\overline{U}_T}\right), \label{eq:lat3a}\\
    \frac{d}{d t}{\overline{L}_m} =& \frac{16j^2l^2b}{3W}\left({A^2_m+B^2_m}+{A^2_T+B^2_T}\right) - l^2A{\overline{L}_m} - \frac{\kappa}{2}\left({\overline{L}_m} -2{\overline{L}_T}\right), \label{eq:lat3b}\\
    \frac{d}{d t}{\overline{U}_T} =& -\frac{4jl^2\lambda^2_R}{3W\left(l^2+\lambda^2_R\right)}\left({A_mB_T-A_TB_m}\right) - l^2A{\overline{U}_T} + \frac{l^2}{l^2+\lambda^2_R}\frac{\kappa}{2}\left({\overline{U}_m}-2{\overline{U}_T}\right)  \label{eq:lat3c}\\
    & + \frac{\lambda^2_R}{l^2+\lambda^2_R}\frac{8RH}{f_0c_p\pi W}, \nonumber\\
    \frac{d}{d t}{\overline{L}_T} =&  \frac{16j^2l^4b}{3W\left(l^2+\lambda_R^2\right)}\left({A_mA_T+B_mB_T}\right) -l^2 A{\overline{L}_{T}} + \frac{l^2}{l^2+\lambda_R^2}\frac{\kappa}{2}\left({\overline{L}_{m}}-2{\overline{L}_{T}}\right) \label{eq:lat3d}\\
    \frac{d}{d t}{A_m}=&-\frac{\alpha_m\beta j}{l}{B_m} +\frac{8\alpha_m j^3(1+b^2)}{3W}\left({B_m\overline{U}_m}+{B_T\overline{U}_T}\right) -\frac{\kappa}{2}\left({A_m}-2{A_T}\right) \label{eq:lat5a}\\
    &-\widetilde{A}_m\alpha_m{A_m} -\frac{8\alpha_mj^2b}{3W}\left({A_m\overline{L}_m}+{A_T\overline{L}_T}\right),
    \nonumber\\
    \frac{d}{d t}{B_m}=& \frac{\alpha_m\beta j}{l}{A_m} -\frac{8\alpha_m j^3(1+b^2)}{3W}\left({A_m\overline{U}_m}+{A_T\overline{U}_T}\right) -\frac{\kappa}{2}\left({B_m}-2{B_T}\right) \label{eq:lat5b}\\
    &-\widetilde{A}_m\alpha_m{B_m} -\frac{8\alpha_mj^2b}{3W}\left({B_m\overline{L}_m}+{B_T\overline{L}_T}\right),\nonumber\\
    \frac{d}{d t}{A_T}=& -\frac{\alpha_T\beta j}{l}{B_T} +\frac{8\alpha_T j^3(1+b^2)}{3W}\left({B_m\overline{U}_T}+{B_T\overline{U}_m}\right) +\frac{\kappa\alpha_T}{2\alpha_m}({A_m}-2{A_T}) \label{eq:lat5c}\\
    &-\widetilde{A}_T\alpha_T{A_T} -\frac{8\alpha_T j \lambda^2_R}{3l^2W}\left({B_m\overline{U}_T}-{B_T\overline{U}_m}\right) -\frac{8\alpha_Tj^2b}{3W}\left({A_m\overline{L}_T}+{A_T\overline{L}_m}\right),\nonumber\\
    \frac{d}{d t}{B_T}=& \frac{\alpha_T\beta j}{l}{A_T} -\frac{8\alpha_T j^3(1+b^2)}{3W}\left({A_m\overline{U}_T}+{A_T\overline{U}_m}\right) +\frac{\kappa\alpha_T}{2\alpha_m}({B_m}-2{B_T}) \label{eq:lat5d}\\
    &-\widetilde{A}_T\alpha_T{B_T} +\frac{8\alpha_T j \lambda^2_R}{3l^2W}\left({A_m\overline{U}_T}-{A_T\overline{U}_m}\right) -\frac{8\alpha_Tj^2b}{3W}\left({B_m\overline{L}_T}+{B_T\overline{L}_m}\right),\nonumber
\end{align}
\end{subequations}
}%
where $\widetilde{A}_m=(j^4l^2+b^4j^4l^2+6b^2j^2l^2+l^2)A$ and $\alpha_m=(j^2+j^2b^2+1)^{-1}$, \newline$\widetilde{A}_T=(j^4l^2+j^4b^4l^2+6j^2b^2l^2+l^2+(j^2+j^2b^2+1)\lambda^2_R)A$ and $\alpha_T=(j^2+j^2b^2+1+\lambda^2_R/l^2)^{-1}$ for better readability.

\section{Purely zonal flow and non-tilted eddies}\label{sec:zonal_flow}
A stationary zonal flow state with the jet located at the centre of the channel and where all eddy components are zero 
is a solution to the equations above: 
\begin{align}
    P^0 = (\overline{U}_m^0,0,\overline{U}_T^0,0,0,0,0,0)= \alpha H(1,0,\frac{l^2 A}{\kappa}+\frac{1}{2},0,0,0,0,0), \label{eq:lat6}
\end{align}
with $\alpha=(A(l^2+\lambda^2_R)(2l^2A+\kappa) +2\kappa l^2A)^{-1}\frac{16\kappa\lambda_R^2R}{f_0c_pl^3W^2}$. Since  eddies are absent, such a solution agrees with the zonal steady state for the  model presented in \cite{Kobras2021Eddy}.  The linear stability of the zonal flow solution can be determined by numerically calculating the eigenvalues of the Jacobian matrix of system (\ref{eq:lat3a})-(\ref{eq:lat5d}). It is extremely challenging to study how the stability of such a state depends on the value of all the parameters of the system. Therefore, while the parameters in table \ref{tab:Fixed_parameters} are kept fixed,  in what follows the heating rate $H$, the eddy streamfunction tilt parameter $b$, and the eddy shape parameter $j$ describing the eddy aspect ratio are varied within the ranges given in table \ref{tab:Varying_parameters}. 

For eddies that are not tilted ($b=0$), this system's stability behaviour is identical to that derived in \cite{Kobras2021Eddy}, as expected. Keeping $j$ fixed, the above-defined purely zonal state grows linearly with the heating $H$. For a further increase of $H$, a complex conjugate pair of eigenvalues of the Jacobian matrix of system (\ref{eq:lat3a})-(\ref{eq:lat5d}) at the steady state (\ref{eq:lat6}) crosses the imaginary axis and their real part becomes positive. \textcolor{black}{It follows that the steady state loses its stability at this threshold value of $H$ and numerical simulations show that in this region of the parameter space a stable periodic limit cycle of the eddy amplitudes $A_{m,T}$ and $B_{m,T}$ appears. This suggests the occurrence of a Hopf bifurcation because only for this type of bifurcation a steady state exchanges stability with a limit cycle. Since these appearing orbits are stable after the switch, the Hopf bifurcation is supercritical.} 
Beyond the bifurcation point, the jet intensity remains constant, whereas the eddy amplitudes ${A_{m,T}}$ and ${B_{m,T}}$ oscillate periodically. 

Apparently, this seems at odds with what is shown in \cite{Kobras2021Eddy}, where the system exhibits a second steady state, the \textit{eddy saturation} regime. Indeed, we are recovering here the same result: the apparent difference results from the fact that in \cite{Kobras2021Eddy} the reduced order model entails the evolution of eddy correlation terms, whereas here we describe the amplitude of the considered eddy components. 

We remark that if $b=0$ the zonally averaged momentum flux (Eq. \ref{eq:mom_flux}) and mixed momentum flux (Eq. \ref{eq:mix_flux}) vanish because non-tilted eddies are symmetric in both directions and therefore the momentum fluxes cancel in the zonal average. 

\section{Barotropic and baroclinic eddy regime}\label{sec:eddy_regimes}

\begin{figure}
\centering
\begin{subfigure}{.99\textwidth}
  \centering
  \includegraphics[width=1\linewidth]{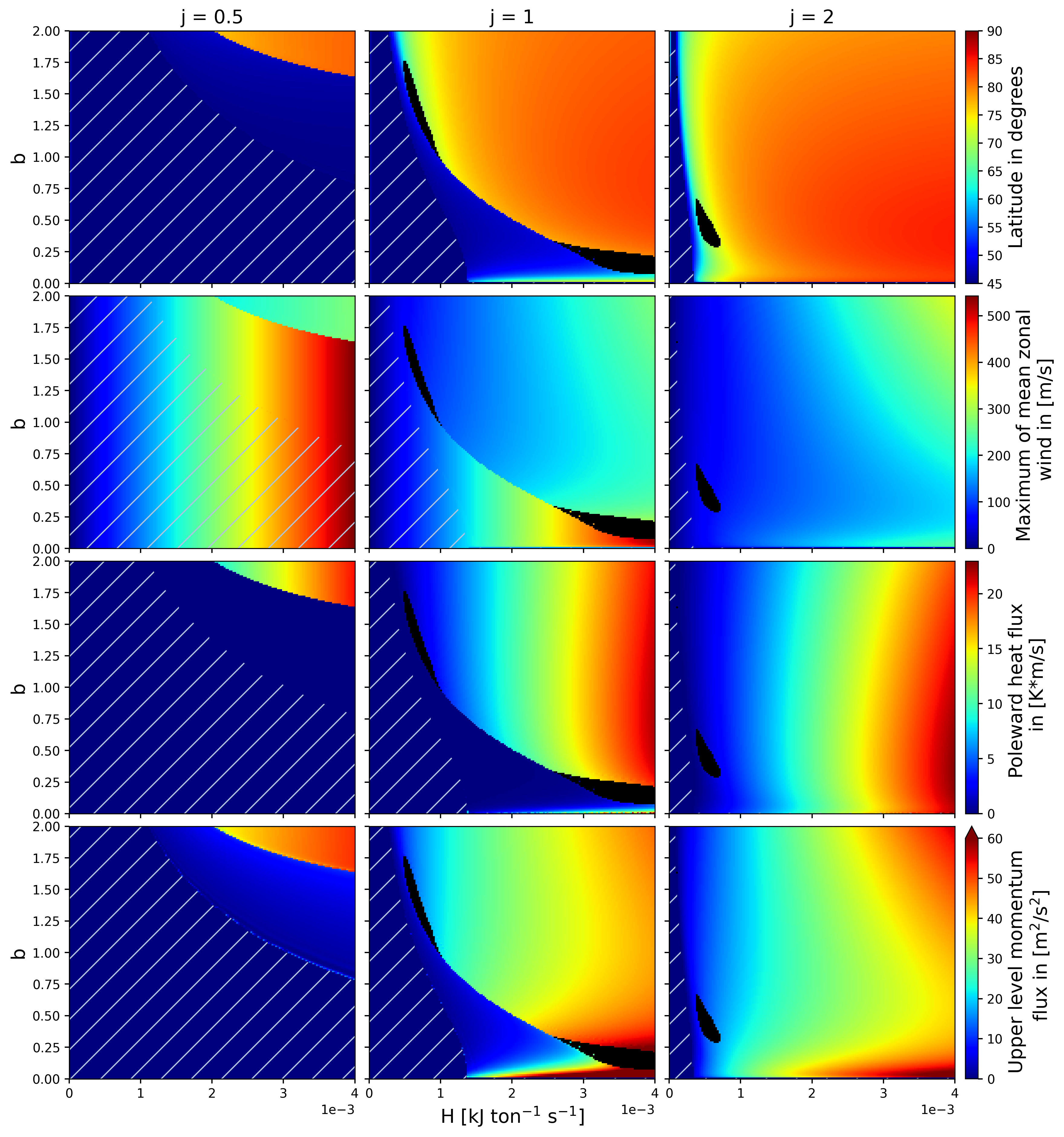}
\end{subfigure}
\caption{Contours of model diagnostics in the H-b-parameter space; for the left column $j=0.5$, the middle $j=1$, the right $j=2$; the first row shows the latitude of the mean zonal wind maximum, the second row shows the maximum values of the mean zonal wind, the third row shows the poleward heat flux, and the fourth row the momentum flux; the dashed areas denote the stable region of the zonal steady state; the black areas are the regions where the \textcolor{black}{zonal flow regime} and baroclinic eddy regimes compete.}
\label{fig:b_H_planes}
\end{figure}

For tilted eddies ($b>0$; see the second and third row of figure \ref{fig:tilted}), the system exhibits an entirely different behaviour compared to the case where eddies are not tilted.  
First of all, the eddy saturation regime disappears: in all cases, the zonal wind speed monotonically increases with $H$. Increasing the tilt leads to a decrease in the critical value of the heating $H$ at which the zonal state loses its stability. Moreover, for larger values of $H$ than this critical value, the dynamics of the system feature two different regimes. 

As before, in both of these unstable regimes the eddy amplitudes ${A_{m,T}}$ and ${B_{m,T}}$ exhibit  periodic behaviour, whereas the mean flow amplitudes ${\overline{U}_{m,T}}$ and ${\overline{L}_{m,T}}$ are non-zero and constant. Therefore, the poleward heat flux and momentum fluxes remain constant as well. \textcolor{black}{The stability of the periodic orbits was determined by numerical experiments. Therefore, the dynamical system was simulated for a dense set of  parameter values within the range shown  in table \ref{tab:Varying_parameters} until asymptotic convergence to constant value for the mean flow amplitudes, poleward heat flux and momentum fluxes. The robustness of the obtained results was verified by simulating the system for a range of different initial conditions, which did not change the final constant values of the solutions. The black areas in figure \ref{fig:b_H_planes} are the regions of the parameter space were the solution did not converge and the behaviour of the model for these parameter combinations is explained later in this section.}

For the following analysis the latitudinal maximum strength and position of the mean zonal wind and shear (Eq. \ref{eq:lat2a}) together with the poleward temperature flux (Eq. \ref{eq:temp_flux}) and the momentum flux on the upper level of the two-level model (Eq. \ref{eq:mom_flux})+(Eq. \ref{eq:mix_flux}) are the quantities of interest. Therefore, we describe the behaviour of these diagnostics directly instead of the evolution of the single amplitudes.

If $b>0$, the maximum of the zonal flow moves poleward when 
the zonal state loses stability. This can be seen in the first row of figure \ref{fig:b_H_planes}, where the latitude of the mean zonal wind maximum is plotted as a function of $H$ and $b$. First, solely the column in the middle is considered, where $j$ was set to one, meaning that the eddy streamfunction has the same zonal and meridional wavenumber, compare figure \ref{fig:tilted}. The dashed region, where the zonal flow peaks in the middle of the channel, 
is the stable region of the zonal steady state.

\begin{figure}[t]
\centering
\begin{subfigure}{.99\textwidth}
  \centering
  \includegraphics[width=1\linewidth]{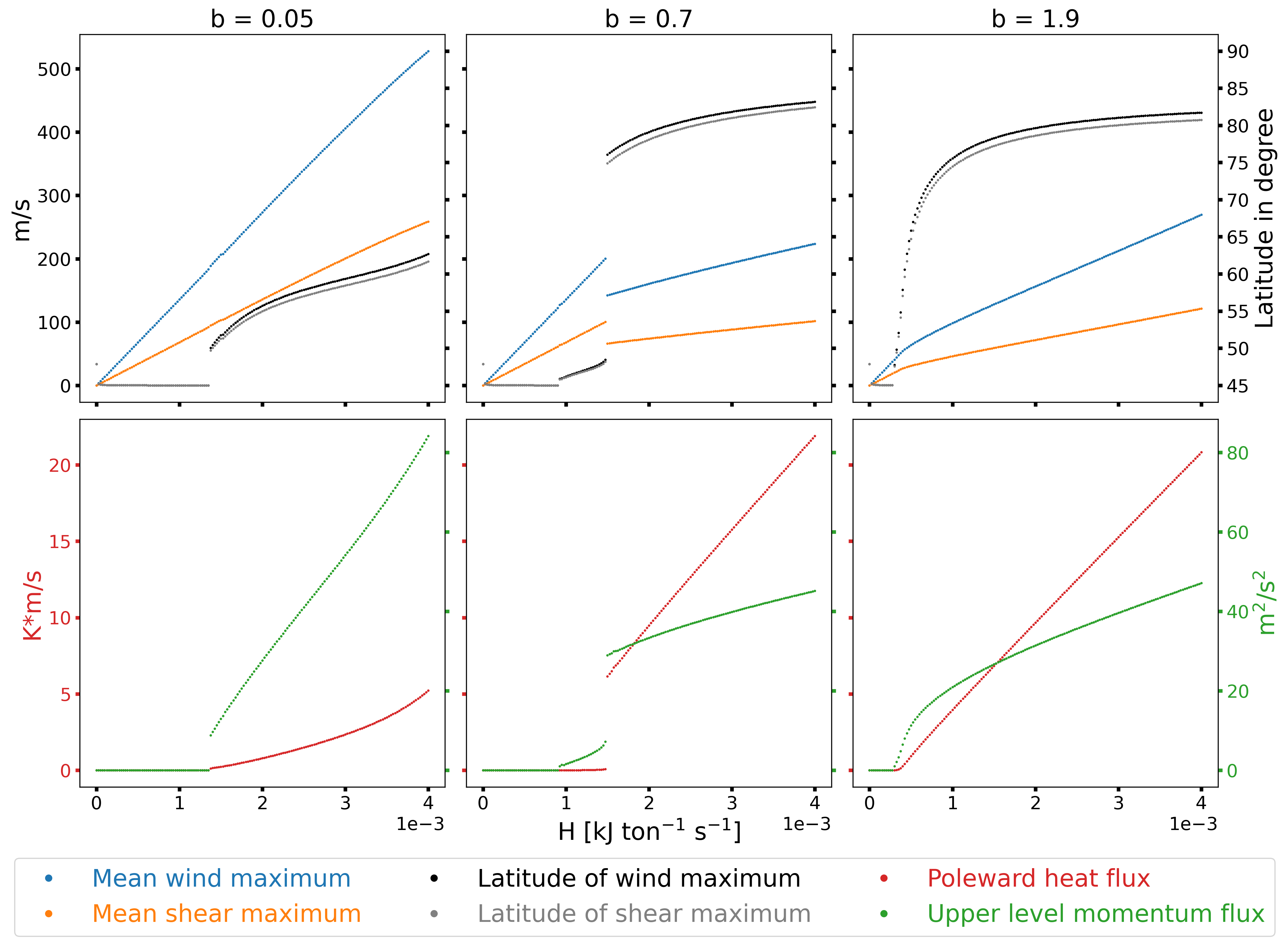}
\end{subfigure}
\caption{Diagnostics of the model as functions of H; the top row plots show the maximum value of the mean zonal wind and shear and their respective latitudinal position, the bottom row shows the poleward heat flux and upper-level momentum flux; for the left plots the tilt $b=0.7$, for the right ones $b=1.9$; for both scenarios the eddy aspect ration $j=1$.}
\label{fig:functions_of_H}
\end{figure}

Increasing the heating $H$ and/or the eddy tilt $b$ leads to a switch of the system to the next state, where the latitudinal maximum of the mean flow is slightly pushed northward and its intensity keeps growing linearly with $H$, see figure \ref{fig:b_H_planes} in the middle of the second row and compare also the top middle plot in figure \ref{fig:functions_of_H}. In this state, the momentum flux (bottom row) is positive and 
increasing with $H$ in a similar manner as the mean flow moves poleward, whereas the poleward heat flux (third row) remains zero.  In the absence of poleward heat flux by the eddies, diffusion is the sole mechanism to balance the equator-to-pole temperature forcing. Hence, the zonal wind speed grows rapidly with $H$. Therefore, this regime is called the barotropic eddy regime in the following. 

For still larger values of $H$ one finds the third regime of the system, where the mean flow is pushed much further northward. In this regime, the zonal wind still grows linearly with the heating, but now with a lower rate, and the poleward heat flux and the momentum flux increase with $H$. Since the heat flux is not zero anymore, this regime is named the baroclinic eddy regime.  \textcolor{black}{An overview of the three flow regimes and the respective flow properties is shown in table \ref{tab:overview_chap4}.}

\begin{table}
\centering
\caption{Overview of flow properties in respective regimes. {\color{black}The expression "with respect to" is shortened as "wrt".}}
\renewcommand*{\arraystretch}{1.1}
\resizebox{\linewidth}{!}{
\begin{tabular}{c|c|c|c}
& Purely zonal & Barotropic & Baroclinic \\
& flow regime & eddy regime & eddy regime  \\
\hline
Mean zonal wind $\overline{u}_m$ & positive, & growth wrt $H$, & reduced growth in $H$\\
and shear $\overline{u}_T$ & growth wrt $H$ & peak pushed north & peak pushed further north\\
\hline
Net poleward heat & 0 & 0 & positive, \\
flux $\overline{v'_m T'_2}$ &  &  & growth wrt $H$ \\
\hline
Eddy momentum & 0 & positive, & positive, \\
flux $\overline{u' v'}$ &  & growth wrt $H$ & reduced growth wrt $H$ \\
\end{tabular}}
\label{tab:overview_chap4}
\end{table}

The switch from the stable state through the barotropic to the baroclinic eddy regime for increasing heating can happen in four different ways, depending on the magnitude of the tilt $b$. Three of these transitions are shown in figure \ref{fig:functions_of_H} for $j=1$.

{\color{black}For $b\leq0.06$ - we portray on the left panels the case $b=0.05$ - one observes for critical value of $H=H_{bc}$ a discontinuous transition from the zonal directly to the baroclinic regime. The transition is accompanied by a modest northward shift of about $5^\circ$ of the peak of the zonal flow, and is followed by a monotonic increase of meridional momentum flux and heat flux with respect to $H$. The momentum flux grows almost linearly with $H$, whilst the heat flux grows, apart from the small offset, as $\approx (H-H_{bc})^{4/3}$, thus featuring increasing sensitivity with larger forcing. The position of the peak of the zonal wind obeys, in the vicinity of the transition, an approximate square root law of the form $(H-H_{bc})^{1/2}$, apart the offset associated with the discontinuity at $H=H_{bc}$. In turn, the intensity of the zonal flow is weakly affected by the presence of the transition, as an approximate linear growth with respect to $H$ is found also in the baroclinic regime, as a result of the weak meridional heat flux.}

For values of $b$ in the interval $[0.34, 0.99]$ (values rounded to two digits) the transition from zonal to baroclinic regime happens in two stages, see panels in the middle ({\color{black}we portray the case $b=0.7$}). First, at a critical value $H=H_{bt}$ the zonal state loses stability in favour of the barotropic regime, leading to growing eddies and therefore an increase in momentum flux, whereas the poleward heat flux remains zero. {\color{black} The transition is accompanied by a very modest ($<1^\circ$) northward shift of the position of the peak of the zonal flow and by a modest increase of the momentum flux. Both of these quantities further increase with respect to $H$.} The second transition from the barotropic into the baroclinic eddy regime {\color{black}occurs at $H=H_{bc}$ and is more obviously discontinuous:} both the momentum and the heat flux jump to a higher value, whereas the zonal flow abruptly slows down and its peak relocated much further north. After the discontinuity,   the momentum and heat flux increase approximately linearly with $H$. The mean flow and the momentum flux continue growing linearly with respect to $H$, but with reduced sensitivities. {\color{black}Note the intensity of the mean flow is oblivious to the transition from the zonal to barotropic regime, because of the lack of a meridional heat transport.}

This regime partly resembles the eddy saturation state found for the non-tilted model. Although the growth rate of the mean flow is not  zero, it is substantially smaller as compared to the other regimes, and the heat flux is growing linearly with the heating. In contrast to the complete eddy saturation with zero mean flow growth rate in the model with non-tilted eddies, here the tilt of the eddies yields a positive poleward momentum flux which increases with $H$ and subsequently generates additional momentum of the mean flow, leading to a small remaining sensitivity of the mean flow to the forcing despite positive poleward heat flux.

The panels on the right hand side of figure \ref{fig:functions_of_H} show the transition behaviour for values of the tilt $b$ higher than $1.76$ - {\color{black}we portray here the case $b=1.9$. Here, as $H$ increases, one finds a continuous transition from the zonal flow regime  through the barotropic eddy regime to the baroclinic eddy regime. In such a regime, one finds a linear growth of the heat flux with respect to $H$, whilst the dependence of the momentum flux with respect to $H$ is $\approx (H-H_{bc})^{1/2}$. The mean zonal wind and shear and the momentum flux show again a reduced growth rate in the baroclinic eddy regime. While the position of the peak of the zonal flow changes continuously with respect to $H$, one finds a very high sensitivity in the proximity of the regime change, with a rapid - with respect to $H$ - northward shift.} 
\begin{figure}[t]
\centering
\begin{subfigure}{.9\textwidth}
  \centering
  \includegraphics[width=1\linewidth]{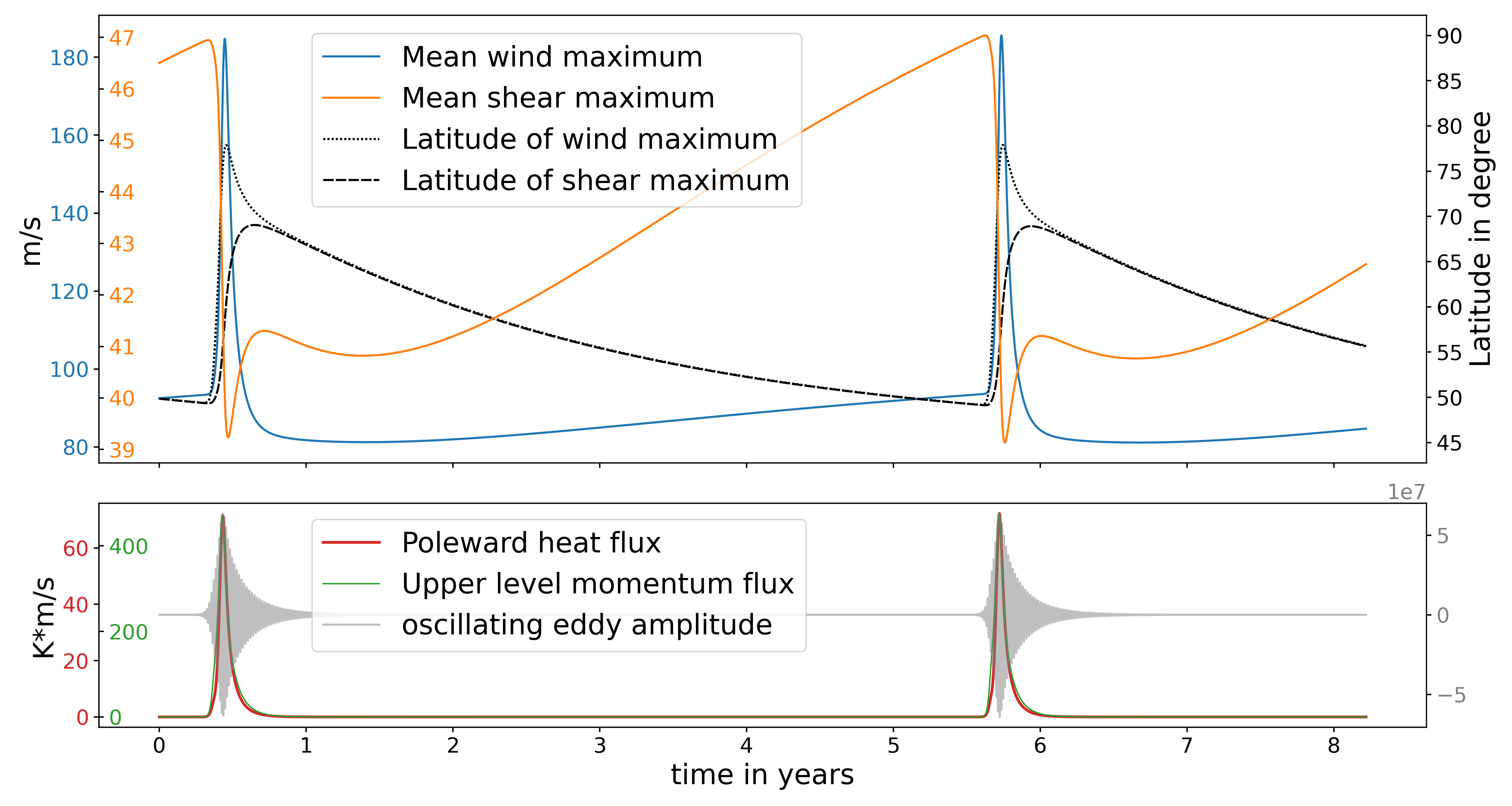}
\end{subfigure}
\caption{Diagnostics of the model as a function of time; $j=1$, $b=1.3$, $H=7.5\times 10^{-4}$, as an example of the recharged oscillator of the model in certain regions of the considered parameter space.}
\label{fig:functions_of_time}
\end{figure}

The fourth possible switch  occurs in the black regions of the heating rate - tilt - plane in figure \ref{fig:b_H_planes}. Here the model exhibits a charge and discharge behaviour where the system oscillates between the purely zonal state and the baroclinic eddy regime on a very long time scale. One cycle of this behaviour for an example set of parameters is shown in figure \ref{fig:functions_of_time}. In the charging phase of the cycle, the mean zonal wind and shear built up and grow over time, building up the baroclinicity in the atmosphere. Starting from further towards the pole, the latitudinal maximum of the flow moves south during this phase almost to the middle of the channel. 

In this phase of the cycle, the eddy amplitudes oscillate in such a way that the heat and momentum fluxes remain zero. After a certain amount of time, the magnitudes of the eddy amplitudes increase rapidly, yielding a sudden peak of the heat and momentum flux magnitudes, which in turn cause the shear and therefore the baroclinicity to collapse, but dramatically increase the mean wind speed and simultaneously push the mean flow far poleward. After this short peak, the fluxes drop again back to zero, the wind speed drops to a lower level than before the peak and the system starts to recharge again. Depending on the tilt $b$, the eddy aspect ratio $j$ and the heating rate $H$, this cycle can vary in the magnitude of the fluctuations as well as in the time scale of the slow variability, ranging from about one year to up to about ten years (not shown).

Finally, the eddy aspect ratio $j$ has a relevant impact on the three different regimes of the model. This can be seen in figure \ref{fig:b_H_planes} in the left and right columns. The left column shows the regimes for $j=0.5$, meaning that the eddies are zonally elongated, see also figure \ref{fig:tilted}, and therefore resemble planetary waves rather than eddies. Such waves have a stabilising effect on the purely zonal flow and also favour the barotropic eddy regime of the model, which can be seen by the increased area of those two regimes compared to the column in the middle. This stabilising effect with respect to baroclinic processes for long waves is well known and results from the $\beta$ effect \cite{Holton2013Dynamic}. 

In contrast to that, in the column on the right, the meridionally elongated eddies for $j=2$ destabilise the purely zonal flow and favour the baroclinic eddy regime, whereas the barotropic eddy regime almost vanishes.  Thus, eddies of this shape, which are also observed in the atmosphere, seem to be more effective in transporting heat poleward.

\section{Summary and Conclusion}\label{sec:discussion}

Starting from the \cite{Phillips1956Thegeneral} classical two-level quasi-geostrophic model on the $\beta$-plane and modifying the derivation and model reduction in \cite{Kobras2021Eddy} we have derived a minimal model for studying the coupling between eddies and zonal flow 
in the mid-latitude atmosphere. This has been accomplished by defining the shape of the mean wind and shear to consist of two modes, allowing for variations in the latitudinal maximum thereof. 
Eddies can differ in terms of their zonal wavenumber and aspect ratio, and the tilt of the streamfunctions can be altered. Depending on these factors, the nature of the eddy-mean flow interaction can change dramatically  \citep{Hoskins1983TheShape}. 
By projecting the quasi-geostrophic equations onto the eddy modes and by considering zonal averages, we have been able to obtain a system of equations describing the coupled evolution of the eddy amplitudes and the mean flow, where the external forcing is given by the  diabatic heating.

If non-tilted eddies are considered, we recover the previous findings presented in \cite{Kobras2021Eddy}. For sufficiently strong baroclinic forcing, the zonal flow loses stability and the eddy saturation regime is reached. In such a state, increases in the strength of the baroclinicity lead to more intense eddies that transport heat efficiently towards the high latitudes, whereas the zonal flow is unaltered. This behaviour is reminiscent of the mechanism of baroclinic adjustment \citep{Stone1978Baroclinic}. 


As soon as we introduce a tilt in the eddies and the zonal flow loses stability, we obtain a 
non-zero momentum flux which pushes the position of the maximum of the jet northward and feeds back eddy kinetic energy into the mean flow. 
The loss of stability of the zonal flow leads the system to growing eddies of finite size. One regime of the flow, which is more readily observed for zonally elongated or circular and rather non-tilted eddies, is dominated by a positive momentum flux increasing with heating and simultaneously speeding up the mean flow. In this regime, the poleward heat flux remains zero. Therefore, this regime is called the barotropic eddy regime.

For meridionally elongated eddies, the system is typically in the baroclinic eddy regime for the considered parameter space. The eddies exhibit a northward positive heat flux which becomes stronger as the baroclinic forcing is strengthened. With intensifying heat imbalance, the momentum flux together with the zonal flow speed increases at a significantly lower rate than in the barotropic regime.

The eddy saturation mechanism is \textit{incomplete}: the strength of the zonal flow has a weaker sensitivity with respect to the strength of the baroclinic forcing than in the case of stable zonal flow, but the adjustment \citep{Stone1978Baroclinic} performed by the eddy heat fluxes is not complete. Such a behaviour, also observed in \cite{Lucarini2007Parametric}, is, in fact, more in agreement with the observations or more comprehensive modelling studies \citep{Novak2018Baroclinic} with respect to the case of full saturation discussed in \cite{Kobras2021Eddy}.

Despite the simplicity of the model and the limiting assumptions made on the eddy streamfunction, especially having a constant tilt, the model resembles the characteristics of the mid-latitude eddies found by \cite{Hoskins1983TheShape} described earlier. Their diagnosed effect of the eddies onto the mean flow is reminiscent of the effect the aspect ratio $j$ in our model has on the two eddy regimes: For \textit{almost} non-tilted, circular eddies, the model favours the barotropic eddy regime with a positive momentum flux enhancing the zonal flow. However, for more zonally elongated southwest-northeast tilted eddies, the model mainly favours the baroclinic regime with reduced mean flow dependency on zonal baroclinic forcing.

Based on a simple oscillator model of baroclinicity (or wind shear) and eddy activity (or heat flux) (\cite{Ambaum2014Anonlinear}), \cite{Novak2015TheLife} used Hoskins' diagnostic to draw a connection between the eddy anisotropy and the three preferred latitudinal positions of the North Atlantic jet stream, which were found by \cite{Woollings2010Variability} analysing North Atlantic winter season ERA-40 data. In agreement with \cite{Franzke2011Persistent}, the study by \cite{Novak2015TheLife} finds the preferred transitions between the south, middle and north regimes of the jet stream. In particular, they also find that enhanced heat fluxes coincide with a preferential northward shift of the jet maximum, as is the case for our minimal model.

It is noteworthy, that the diagnostic quantities by \cite{Hoskins1983TheShape} and \cite{Novak2015TheLife} are averaged in time, whereas the model in the present paper deals with a zonally averaged flow without variation in the eddy shape. Therefore, instead of an evolution of the flow in time, the barotropic and baroclinic regimes of our model can rather be interpreted as a snapshot in time of the life cycle of the eddies in the mid-latitude storm track.

\textcolor{black}{Further to that, the results and conclusions, such as the extent of the eddy saturation or the sensitivity of the jets and eddy fluxes to various internal and external parameters, are all in the context of the simple model derived here. The same applies to the potentially interesting power law scalings found for various model diagnostics in the vicinity of the critical transitions of the model. The robustness of these results is difficult to assess a priori, since there are many different ways in which the models can be modified and/or extended in the direction of higher complexity. To show such a robustness, one would need to consider a hierarchy of models with an increasing number of modes and analyse systematically their behaviour. It is encouraging that, as mentioned above, our model features the eddy saturation mechanism, which is observed in comprehensive atmospheric and oceanic models. Clearly, further work is needed in this direction and is left for future investigations.} 
Concluding, this study presents a minimal model of the mid-latitude atmosphere exhibiting eddy-mean flow interaction, jet shifts, and partial eddy saturation. Despite the minimal setup, our model captures key processes of the mid-latitude atmosphere on different time scales, which are commonly observed in re-analysis data and atmospheric climate models.  
 Melanie Kobras is supported by the U.K. Engineering and Physical Sciences Research Council (Grant EP/R513301/1). Valerio Lucarini acknowledges the support received from the EPSRC project EP/T018178/1, from the EU Horizon 2020 project Tipping Points in the Earth System (TiPES) (Grant no. 820970), and from the Marie Curie ITN CriticalEarth (Grant no. 956170)

\newpage
\appendix

\section{Parameter space}\label{apx:Parameter_space}
\newcommand{\Times}{\!\!\times\!\!}
\begin{table}[h]
\centering
\caption{Fixed parameter values for stability analysis.}
\renewcommand*{\arraystretch}{1.3}
\resizebox{\linewidth}{!}{
\begin{tabular}{lcccccccc}\hline
Parameter & $\lambda^2_{R}$ & $\beta$ & $A$ & $\kappa$ & $R$ & $f_0$ & $c_{p}$ & $W$ \\
\hline
Value & $4.39\Times10^{-12}$ & $1.6\Times10^{-11}$ & $10^5$ & $4\Times10^{-6}$ & $287$ & $10^{-4}$ & $1004$ & $10^{7}$ \\
Unit & $\text{m}^{-2}$ & $\text{m}^{-1}\text{s}^{-1}$ & $\text{m}^{2}\text{s}^{-1}$ & $\text{s}^{-1}$ & $\text{JK}^{-1}\text{kg}^{-1}$ & $\text{s}^{-1}$ & $\text{JK}^{-1}\text{kg}^{-1}$ & $\text{m}$ \\
\hline
\end{tabular}}
\label{tab:Fixed_parameters}
\end{table}

\begin{table}[h]
\centering
\caption{Varying parameter values for stability analysis.}
\renewcommand*{\arraystretch}{1.1}
\begin{tabular}{lccc}\hline
Parameter
& $H$ & $b$ & $j$ \\
\hline
Range & $(0, 4\Times10^{-3}]$ & $[0, 2]$ & $0.5, 1, 2$ \\
Unit & $\text{KJ ton}^{-1}\text{ s}^{-1}$ & - & -\\
\hline
\end{tabular}
\label{tab:Varying_parameters}
\end{table}


\begin{thebibliography}{25}
\expandafter\ifx\csname natexlab\endcsname\relax\def\natexlab#1{#1}\fi
\providecommand{\url}[1]{\texttt{#1}}
\providecommand{\href}[2]{#2}
\providecommand{\path}[1]{#1}
\providecommand{\DOIprefix}{doi:}
\providecommand{\ArXivprefix}{arXiv:}
\providecommand{\URLprefix}{URL: }
\providecommand{\Pubmedprefix}{pmid:}
\providecommand{\doi}[1]{\href{http://dx.doi.org/#1}{\path{#1}}}
\providecommand{\Pubmed}[1]{\href{pmid:#1}{\path{#1}}}
\providecommand{\bibinfo}[2]{#2}
\ifx\xfnm\relax \def\xfnm[#1]{\unskip,\space#1}\fi
\bibitem[{Ambaum and Novak(2014)}]{Ambaum2014Anonlinear}
\bibinfo{author}{Ambaum, M.H.P.}, \bibinfo{author}{Novak, L.},
  \bibinfo{year}{2014}.
\newblock \bibinfo{title}{A nonlinear oscillator describing storm track
  variability}.
\newblock \bibinfo{journal}{Quarterly Journal of the Royal Meteorological
  Society} \bibinfo{volume}{140}, \bibinfo{pages}{2680--2684}.
\newblock \DOIprefix\doi{10.1002/qj.2352}.
\bibitem[{Butler et~al.(2010)Butler, Thompson and Heikes}]{Butler2010}
\bibinfo{author}{Butler, A.H.}, \bibinfo{author}{Thompson, D.W.J.},
  \bibinfo{author}{Heikes, R.}, \bibinfo{year}{2010}.
\newblock \bibinfo{title}{The steady-state atmospheric circulation response to
  climate change-like thermal forcings in a simple general circulation
  model}.
\newblock \bibinfo{journal}{Journal of Climate} \bibinfo{volume}{23},
  \bibinfo{pages}{3474 -- 3496}.
\newblock \URLprefix
  \url{https://journals.ametsoc.org/view/journals/clim/23/13/2010jcli3228.1.xml},
  \DOIprefix\doi{https://doi.org/10.1175/2010JCLI3228.1}.
\bibitem[{Franzke et~al.(2011)Franzke, Woollings and
  Martius}]{Franzke2011Persistent}
\bibinfo{author}{Franzke, C.}, \bibinfo{author}{Woollings, T.},
  \bibinfo{author}{Martius, O.}, \bibinfo{year}{2011}.
\newblock \bibinfo{title}{Persistent circulation regimes and preferred regime
  transitions in the north atlantic}.
\newblock \bibinfo{journal}{Journal of the Atmospheric Sciences}
  \bibinfo{volume}{68}, \bibinfo{pages}{2809--2825}.
\newblock \DOIprefix\doi{10.1175/jas-d-11-046.1}.
\bibitem[{Frierson(2008)}]{Frierson2008}
\bibinfo{author}{Frierson, D.M.W.}, \bibinfo{year}{2008}.
\newblock \bibinfo{title}{Midlatitude static stability in simple and
  comprehensive general circulation models}.
\newblock \bibinfo{journal}{Journal of the Atmospheric Sciences}
  \bibinfo{volume}{65}, \bibinfo{pages}{1049 -- 1062}.
\newblock \URLprefix
  \url{https://journals.ametsoc.org/view/journals/atsc/65/3/2007jas2373.1.xml},
  \DOIprefix\doi{https://doi.org/10.1175/2007JAS2373.1}.
\bibitem[{Fyfe(2003)}]{Fyfe2003Extratropical}
\bibinfo{author}{Fyfe, J.C.}, \bibinfo{year}{2003}.
\newblock \bibinfo{title}{Extratropical southern hemisphere cyclones:
  Harbingers of climate change?}
\newblock \bibinfo{journal}{Journal of Climate} \bibinfo{volume}{16},
  \bibinfo{pages}{2802--2805}.
\newblock \DOIprefix\doi{10.1175/1520-0442(2003)016<2802:eshcho>2.0.co;2}.
\bibitem[{Holton and Hakim(2013)}]{Holton2013Dynamic}
\bibinfo{author}{Holton, J.R.}, \bibinfo{author}{Hakim, G.J.},
  \bibinfo{year}{2013}.
\newblock \bibinfo{title}{An Introduction to Dynamic Meteorology}.
\newblock \bibinfo{edition}{5} ed., \bibinfo{publisher}{Academic Press}.
\bibitem[{Hoskins et~al.(1983)Hoskins, James and White}]{Hoskins1983TheShape}
\bibinfo{author}{Hoskins, B.J.}, \bibinfo{author}{James, I.N.},
  \bibinfo{author}{White, G.H.}, \bibinfo{year}{1983}.
\newblock \bibinfo{title}{The shape, propagation and mean-flow interaction of
  large-scale weather systems}.
\newblock \bibinfo{journal}{Journal of the Atmospheric Sciences}
  \bibinfo{volume}{40}, \bibinfo{pages}{1595--1612}.
\newblock \DOIprefix\doi{10.1175/1520-0469(1983)040<1595:tspamf>2.0.co;2}.
\bibitem[{K{\aa}llberg et~al.(2005)K{\aa}llberg, Berrisford, Hoskins, Simmons,
  Uppala, Lamy-Thepaut and Hine}]{Kallberg2005ERA-40}
\bibinfo{author}{K{\aa}llberg, P.}, \bibinfo{author}{Berrisford, P.},
  \bibinfo{author}{Hoskins, B.}, \bibinfo{author}{Simmons, A.},
  \bibinfo{author}{Uppala, S.}, \bibinfo{author}{Lamy-Thépaut, S.},
  \bibinfo{author}{Hine, R.}, \bibinfo{year}{2005}.
\newblock \bibinfo{title}{ERA-40 Atlas}.
\newblock Number 19, \bibinfo{publisher}{ECMWF}.
\newblock \URLprefix
  \url{https://sites.ecmwf.int/era/40-atlas/docs/index.html}.
\bibitem[{Kidston et~al.(2010)Kidston, Frierson, Renwick and
  Vallis}]{Kidson2010}
\bibinfo{author}{Kidston, J.}, \bibinfo{author}{Frierson, D.M.W.},
  \bibinfo{author}{Renwick, J.A.}, \bibinfo{author}{Vallis, G.K.},
  \bibinfo{year}{2010}.
\newblock \bibinfo{title}{Observations, simulations, and dynamics of jet stream
  variability and annular modes}.
\newblock \bibinfo{journal}{Journal of Climate} \bibinfo{volume}{23},
  \bibinfo{pages}{6186 -- 6199}.
\newblock \URLprefix
  \url{https://journals.ametsoc.org/view/journals/clim/23/23/2010jcli3235.1.xml},
  \DOIprefix\doi{https://doi.org/10.1175/2010JCLI3235.1}.
\bibitem[{Kobras et~al.(2021)Kobras, Ambaum and Lucarini}]{Kobras2021Eddy}
\bibinfo{author}{Kobras, M.}, \bibinfo{author}{Ambaum, M.H.P.},
  \bibinfo{author}{Lucarini, V.}, \bibinfo{year}{2021}.
\newblock \bibinfo{title}{Eddy saturation in a reduced two-level model of the
  atmosphere}.
\newblock \bibinfo{journal}{Geophysical \& Astrophysical Fluid Dynamics} ,
  \bibinfo{pages}{1--18}\DOIprefix\doi{10.1080/03091929.2021.1990912}.
\bibitem[{Koo and Ghil(2002)}]{Koo2002}
\bibinfo{author}{Koo, S.}, \bibinfo{author}{Ghil, M.}, \bibinfo{year}{2002}.
\newblock \bibinfo{title}{Successive bifurcations in a simple model of
  atmospheric zonal-flow vacillation}.
\newblock \bibinfo{journal}{Chaos: An Interdisciplinary Journal of Nonlinear
  Science} \bibinfo{volume}{12}, \bibinfo{pages}{300--309}.
\newblock \URLprefix \url{https://doi.org/10.1063/1.1468249},
  \DOIprefix\doi{10.1063/1.1468249},
  \href{http://arxiv.org/abs/https://doi.org/10.1063/1.1468249}{{\tt
  arXiv:https://doi.org/10.1063/1.1468249}}.
\bibitem[{Lorenz and DeWeaver(2007)}]{Lorenz2007}
\bibinfo{author}{Lorenz, D.J.}, \bibinfo{author}{DeWeaver, E.T.},
  \bibinfo{year}{2007}.
\newblock \bibinfo{title}{Tropopause height and zonal wind response to global
  warming in the ipcc scenario integrations}.
\newblock \bibinfo{journal}{Journal of Geophysical Research: Atmospheres}
  \bibinfo{volume}{112}.
\newblock \URLprefix
  \url{https://agupubs.onlinelibrary.wiley.com/doi/abs/10.1029/2006JD008087},
  \DOIprefix\doi{https://doi.org/10.1029/2006JD008087},
  \href{http://arxiv.org/abs/https://agupubs.onlinelibrary.wiley.com/doi/pdf/10.1029/2006JD008087}{{\tt
  arXiv:https://agupubs.onlinelibrary.wiley.com/doi/pdf/10.1029/2006JD008087}}.
\bibitem[{Lorenz(1955)}]{Lorenz1955Available}
\bibinfo{author}{Lorenz, E.N.}, \bibinfo{year}{1955}.
\newblock \bibinfo{title}{Available potential energy and the maintenance of the
  general circulation}.
\newblock \bibinfo{journal}{Tellus} \bibinfo{volume}{7},
  \bibinfo{pages}{157--167}.
\newblock \DOIprefix\doi{10.3402/tellusa.v7i2.8796}.
\bibitem[{Lorenz(1967)}]{Lorenz1967Thenature}
\bibinfo{author}{Lorenz, E.N.}, \bibinfo{year}{1967}.
\newblock \bibinfo{title}{The nature and theory of the general circulation of
  the atmosphere}. volume \bibinfo{volume}{218}.
\newblock \bibinfo{publisher}{World Meteorological Organization Geneva}.
\bibitem[{Lu et~al.(2010)Lu, Chen and Frierson}]{Lu2010}
\bibinfo{author}{Lu, J.}, \bibinfo{author}{Chen, G.},
  \bibinfo{author}{Frierson, D.M.W.}, \bibinfo{year}{2010}.
\newblock \bibinfo{title}{The position of the midlatitude storm track and
  eddy-driven westerlies in aquaplanet agcms}.
\newblock \bibinfo{journal}{Journal of the Atmospheric Sciences}
  \bibinfo{volume}{67}, \bibinfo{pages}{3984 -- 4000}.
\newblock \URLprefix
  \url{https://journals.ametsoc.org/view/journals/atsc/67/12/2010jas3477.1.xml},
  \DOIprefix\doi{https://doi.org/10.1175/2010JAS3477.1}.
\bibitem[{Lucarini et~al.(2007)Lucarini, Speranza and
  Vitolo}]{Lucarini2007Parametric}
\bibinfo{author}{Lucarini, V.}, \bibinfo{author}{Speranza, A.},
  \bibinfo{author}{Vitolo, R.}, \bibinfo{year}{2007}.
\newblock \bibinfo{title}{Parametric smoothness and self-scaling of the
  statistical properties of a minimal climate model: What beyond the mean field
  theories?}
\newblock \bibinfo{journal}{Physica D: Nonlinear Phenomena}
  \bibinfo{volume}{234}, \bibinfo{pages}{105--123}.
\newblock \DOIprefix\doi{10.1016/j.physd.2007.07.006}.
\bibitem[{Novak et~al.(2018)Novak, Ambaum and Harvey}]{Novak2018Baroclinic}
\bibinfo{author}{Novak, L.}, \bibinfo{author}{Ambaum, M.H.P.},
  \bibinfo{author}{Harvey, B.J.}, \bibinfo{year}{2018}.
\newblock \bibinfo{title}{Baroclinic adjustment and dissipative control of
  storm tracks}.
\newblock \bibinfo{journal}{Journal of the Atmospheric Sciences}
  \bibinfo{volume}{75}, \bibinfo{pages}{2955--2970}.
\newblock \DOIprefix\doi{10.1175/jas-d-17-0210.1}.
\bibitem[{Novak et~al.(2015)Novak, Ambaum and Tailleux}]{Novak2015TheLife}
\bibinfo{author}{Novak, L.}, \bibinfo{author}{Ambaum, M.H.P.},
  \bibinfo{author}{Tailleux, R.}, \bibinfo{year}{2015}.
\newblock \bibinfo{title}{The life cycle of the north atlantic storm track}.
\newblock \bibinfo{journal}{Journal of the Atmospheric Sciences}
  \bibinfo{volume}{72}, \bibinfo{pages}{821--833}.
\newblock \DOIprefix\doi{10.1175/jas-d-14-0082.1}.
\bibitem[{Orlanski(1998)}]{Orlanski1998Poleward}
\bibinfo{author}{Orlanski, I.}, \bibinfo{year}{1998}.
\newblock \bibinfo{title}{Poleward deflection of storm tracks}.
\newblock \bibinfo{journal}{Journal of the Atmospheric Sciences}
  \bibinfo{volume}{55}, \bibinfo{pages}{2577--2602}.
\newblock \DOIprefix\doi{10.1175/1520-0469(1998)055<2577:pdost>2.0.co;2}.
\bibitem[{Phillips(1956)}]{Phillips1956Thegeneral}
\bibinfo{author}{Phillips, N.A.}, \bibinfo{year}{1956}.
\newblock \bibinfo{title}{The general circulation of the atmosphere: A
  numerical experiment}.
\newblock \bibinfo{journal}{Quarterly Journal of the Royal Meteorological
  Society} \bibinfo{volume}{82}, \bibinfo{pages}{123--164}.
\newblock \DOIprefix\doi{10.1002/qj.49708235202}.
\bibitem[{Stone(1978)}]{Stone1978Baroclinic}
\bibinfo{author}{Stone, P.H.}, \bibinfo{year}{1978}.
\newblock \bibinfo{title}{Baroclinic adjustment}.
\newblock \bibinfo{journal}{Journal of Atmospheric Sciences}
  \bibinfo{volume}{35}, \bibinfo{pages}{561--571}.
\newblock \DOIprefix\doi{10.1175/1520-0469(1978)035<0561:Ba>2.0.Co;2}.
\bibitem[{Thompson and Wallace(2000)}]{Thompson2000}
\bibinfo{author}{Thompson, D.W.J.}, \bibinfo{author}{Wallace, J.M.},
  \bibinfo{year}{2000}.
\newblock \bibinfo{title}{Annular modes in the extratropical circulation. part
  i: Month-to-month variability}.
\newblock \bibinfo{journal}{Journal of Climate} \bibinfo{volume}{13},
  \bibinfo{pages}{1000 -- 1016}.
\newblock \URLprefix
  \url{https://journals.ametsoc.org/view/journals/clim/13/5/1520-0442_2000_013_1000_amitec_2.0.co_2.xml},
  \DOIprefix\doi{https://doi.org/10.1175/1520-0442(2000)013<1000:AMITEC>2.0.CO;2}.
\bibitem[{Thompson(1987)}]{Thompson1987}
\bibinfo{author}{Thompson, P.D.}, \bibinfo{year}{1987}.
\newblock \bibinfo{title}{Large-scale dynamical response to differential
  heating: Statistical equilibrium states and amplitude vacillation}.
\newblock \bibinfo{journal}{Journal of Atmospheric Sciences}
  \bibinfo{volume}{44}, \bibinfo{pages}{1237 -- 1248}.
\newblock \URLprefix
  \url{https://journals.ametsoc.org/view/journals/atsc/44/8/1520-0469_1987_044_1237_lsdrtd_2_0_co_2.xml},
  \DOIprefix\doi{https://doi.org/10.1175/1520-0469(1987)044<1237:LSDRTD>2.0.CO;2}.
\bibitem[{Woollings et~al.(2010)Woollings, Hannachi and
  Hoskins}]{Woollings2010Variability}
\bibinfo{author}{Woollings, T.}, \bibinfo{author}{Hannachi, A.},
  \bibinfo{author}{Hoskins, B.}, \bibinfo{year}{2010}.
\newblock \bibinfo{title}{Variability of the north atlantic eddy-driven jet
  stream}.
\newblock \bibinfo{journal}{Quarterly Journal of the Royal Meteorological
  Society} \bibinfo{volume}{136}, \bibinfo{pages}{856--868}.
\newblock \DOIprefix\doi{10.1002/qj.625}.
\bibitem[{Yin(2005)}]{Yin2005Aconstistant}
\bibinfo{author}{Yin, J.H.}, \bibinfo{year}{2005}.
\newblock \bibinfo{title}{A consistent poleward shift of the storm tracks in
  simulations of 21st century climate}.
\newblock \bibinfo{journal}{Geophysical Research Letters} \bibinfo{volume}{32}.
\newblock \DOIprefix\doi{10.1029/2005gl023684}.

\end{thebibliography}
\end{document}